\documentclass[aip,twocolumn,showpacs,amsmath,amssymb,preprintnumbers,superscriptaddress,10pt,longbibliography]{revtex4-1}

\usepackage{graphicx}
\usepackage{SIunits}
\usepackage{hyphenat}
\usepackage[bookmarks=false]{hyperref}
\usepackage{color}
\usepackage[capitalise]{cleveref}
\crefname{section}{Sec.}{Secs.}
\Crefname{section}{Section}{Sections}
\usepackage{physics}

\definecolor{pink}{RGB}{255,0,255}

\definecolor{red}{rgb}{0,0,1}

\definecolor{darkgreen}{RGB}{0,130,0}

\definecolor{orange}{RGB}{255,140,0}
\definecolor{orange1}{RGB}{200,50,0}

\begin{document}
	
\title{\Large{Faking photon number on a transition-edge sensor}}

\author{Poompong~Chaiwongkhot}
\email{poompong.ch@gmail.com}
\affiliation{Institute for Quantum Computing, University of Waterloo, Waterloo, ON, N2L~3G1 Canada}
\affiliation{Department of Physics and Astronomy, University of Waterloo, Waterloo, ON, N2L~3G1 Canada}
\affiliation{Department of Physics, Faculty of Science, Mahidol University, Bangkok, 10400 Thailand}
\affiliation{Quantum technology foundation (Thailand), Bangkok, 10110 Thailand}

\author{Jiaqiang~Zhong}
\affiliation{Purple Mountain Observatory and Key Laboratory of Radio Astronomy, Chinese Academy of Sciences, 10~Yuanhua road, Nanjing 210033, People's Republic of China}

\author{Anqi~Huang}
\email{angelhuang.hn@gmail.com}
\affiliation{Institute for Quantum Information \& State Key Laboratory of High Performance Computing, College of Computer, National University of Defense Technology, Changsha 410073, People's Republic of China}

\author{Hao~Qin}
\affiliation{CAS Quantum Network Co.,\ Ltd.,\ 99 Xiupu road, Shanghai 201315, People's Republic of China}

\author{Sheng-cai~Shi}
\affiliation{Purple Mountain Observatory and Key Laboratory of Radio Astronomy, Chinese Academy of Sciences, 10~Yuanhua road, Nanjing 210033, People's Republic of China}

\author{Vadim~Makarov}
\affiliation{Russian Quantum Center, Skolkovo, Moscow 121205, Russia}
\affiliation{Shanghai Branch, National Laboratory for Physical Sciences at Microscale and CAS Center for Excellence in Quantum Information, University of Science and Technology of China, Shanghai 201315, People's Republic of China}
\affiliation{NTI Center for Quantum Communications, National University of Science and Technology MISiS, Moscow 119049, Russia}

\date{\today}

\begin{abstract}
We study potential security vulnerabilities of a single-photon detector based on superconducting transition-edge sensor. In a simple experiment, we show that an adversary could fake a photon number result at a certain wavelength by sending a larger number of photons at a longer wavelength. In another experiment, we show that the detector can be blinded by bright continuous-wave light and then, a controlled response simulating single-photon detection can be produced by applying a bright light pulse. We model an intercept-and-resend attack on a quantum key distribution system that exploits the latter vulnerability and, under certain assumptions, succeeds to steal the key.
\end{abstract}

\maketitle


Photon detectors are indispensable in quantum communication applications \cite{hadfield2009}. To ensure the reliability of detection results, it is important to characterize the detectors being used both within the intended working parameters and possible unintended conditions. This characterization could help in revealing possible flaws and imperfections. These flaws could lead to misguided detection results or, worse, exploitable vulnerabilities in the case of quantum cryptography applications. This characterization guides the work on improving the robustness of quantum systems. Over the years, many attacks have been reported on various types of photon detectors based on avalanche photodiodes \cite{makarov2006,zhao2008,lydersen2010a,lydersen2010,lydersen2010b,lydersen2011,gerhardt2011,huang2016,qian2018,fei2018} and superconducting nanowires \cite{lydersen2011c,tanner2014,elezov2019}.

Transition-edge sensor (TES) is a photon detector capable of providing full photon-number-resolving capability \cite{berggren2013,eisaman2011}. It also achieved the highest detection efficiency among photon-number-resolving detectors up to $95\%$ at $1550~\nano\meter$ \cite{lita2008,fukuda2009,miller2011b}. This type of detector is used in various applications that require high detection probability, such as loophole-free Bell test \cite{giustina2015}. Its photon number resolving capability could also be used to monitor against attacks on a quantum key distribution (QKD) system \cite{xu2010e}. As one of the potential detectors in quantum communication where the reliability of detection result affects overall security, the TES photon detector should be investigated for its robustness and possible flaws. In this study, we experimentally demonstrate two potential vulnerabilities of TES, namely, a wavelength attack where the photon number result could be controlled by changing signal's wavelength and a faked-state attack where the adversary increases the temperature of TES with an appropriate bright continuous-wave (CW) laser then forces an arbitrary photon number detection result using a bright pulsed laser. 



\begin{figure}
	\includegraphics{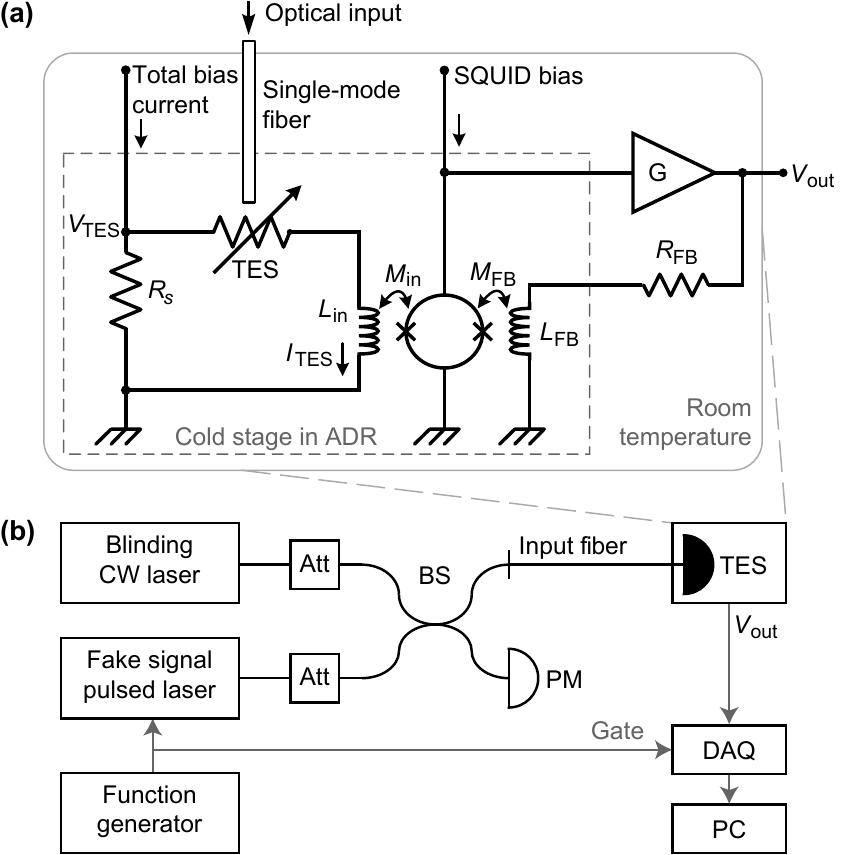}
	\caption{Experimental setup. (a)~Internal circuit diagram of the TES system, consisting of the TES photon detector and its DC-SQUID readout. The TES photon detector is mounted on a $100$-$\milli\kelvin$ cold stage chilled by an adiabatic demagnetization refrigerator (ADR). The TES current $I_{\text{TES}}$ is readout by DC-SQUID electronics and transferred proportionally to a voltage output $V_{\text{out}}$. (b)~Blinding and fake signal power is controlled by variable attenuators (Att), combined at a $50\!:\!50$ fiber-optic beam splitter (BS), measured by an optical power meter (PM), and applied to the TES system under test. Its output voltage $V_{\text{out}}$ is recorded and analyzed by a data acquisition module (DAQ) connected to a computer (PC).}
	\label{fig:setup}
\end{figure}

A transition-edge sensor is a sensitive micro-calorimeter whose sensing element consists of an absorber and a superconductive thermometer with a positive temperature coefficient of resistance ($\dd{R}/\dd{T}>0$) \cite{irwin2005}. During the operation, the sensing element's temperature is kept near the transition temperature via voltage-biasing \cite{irwin_application_1995}. This voltage-biasing is provided by an external total bias current flowing through a shunt resistor $R_s$ connected in parallel with the TES [\cref{fig:setup}(a)]. In our setup $R_s = 16.1~\milli\ohm$, which is much smaller than the TES normal-conductivity resistance of $3~\ohm$. 

The current passing through the TES $I_{\text{TES}}$ flows through an inductive coil $L_\text{in}$. The latter couples its magnetic flux via a mutual inductance ($M_\text{in}$)  to a direct-current superconducting quantum interference device (DC-SQUID). The SQUID serves as a low-noise amplifier of $I_{\text{TES}}$. A feedback coil $L_\text{FB}$ inside the ADR, together with a room-temperature amplifier G and feedback resistor $R_\text{FB}$ are used to transform the signal from the TES into a measurable voltage $V_\text{out}$ \cite{drung2006}. $I_{\text{TES}}$ is obtained by dividing $V_\text{out}$ by the current-to-voltage gain of the DC-SQUID and amplifier G ($0.375~\volt\per\micro\ampere$ in this experiment), while the voltage across TES $V_\text{TES}$ is calculated by multiplying $R_s$ by the current through it (total bias current with $I_{\text{TES}}$ subtracted). 

When a photon from the input optical fiber hits the detector, the photon's energy is absorbed, raising the TES' temperature and resistance. This change of resistance reduces $I_{\text{TES}}$ and proportionally reduces $V_{\text{out}}$. From the relation of TES temperature and $I_{\text{TES}}$, it can be seen that the change of $V_\text{out}$ during the detection is proportional to the absorbed energy of the photon(s), enabling photon-number discrimination.

In our setup, the TES and SQUID are attached on a copper block attached in turn to the cold plate of the ADR. Under normal operating conditions, both the TES and SQUID are at $100~\milli\kelvin$ temperature. Their bias currents are provided by specialised electronic circuits (commercially available from Magnicon GmbH). 

\begin{figure}
	\includegraphics{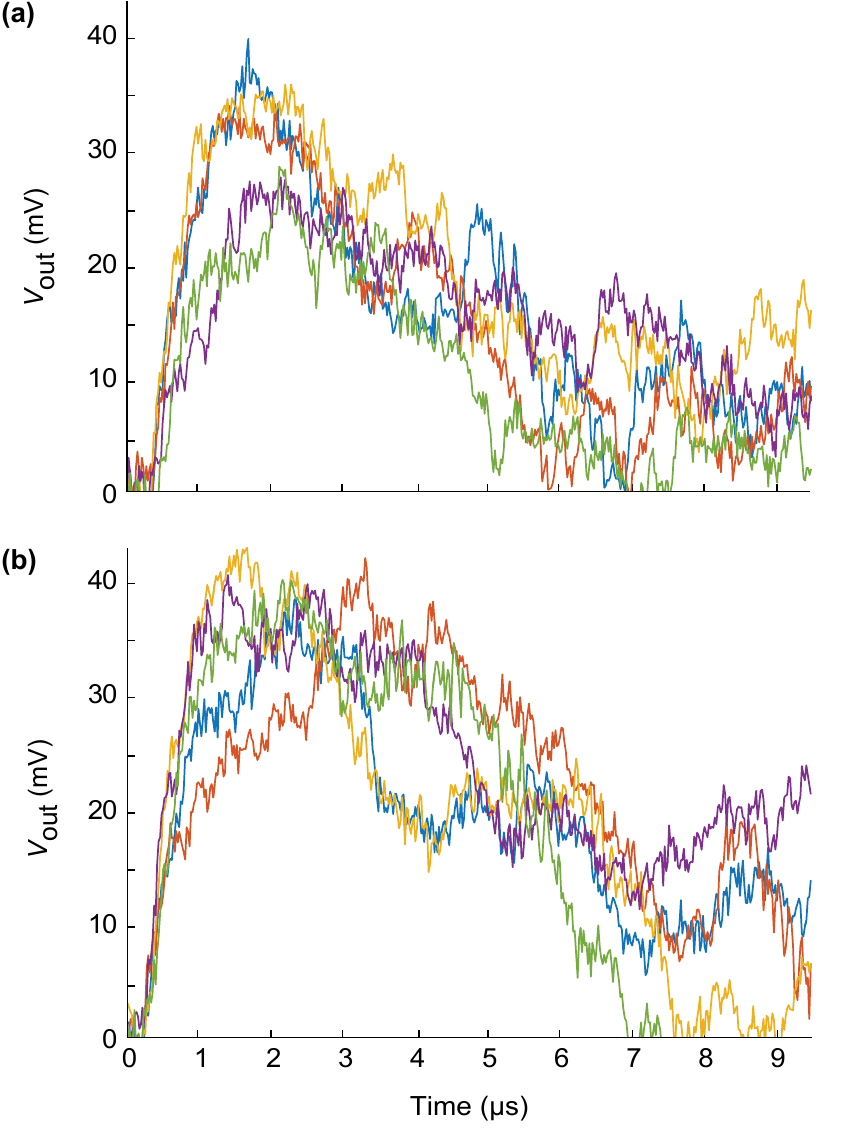}
	\caption{Oscillograms of $V_\text{out}$. (a)~Typical single-photon responses. (b)~Fake ``single-photon'' responses under $1550~\nano\meter$ blinding attack with $2.4\times10^{-18}~\joule$ pulse energy (i.e.,\ about 19-photon weak coherent pulse).}
	\label{fig:oscillograms}
\end{figure}

To measure the response of the TES to various optical signals, we use a setup shown in \cref{fig:setup}(b). The TES is a fiber-coupled $10\times10~\micro\meter$ Ti device in a multilayer optical resonator designed to maximise coupling at $1550~\nano\meter$ wavelength and is similar to devices reported in \cite{fukuda2009,fukuda2011}. The photon coupling efficiency in our TES sample under test is $\approx1\%$ owing to a misaligned fiber end to the TES effective area. However, this should not affect the results of our study in a qualitative way, because the misalignment merely introduces additional optical attenuation and can be compensated by applying a brighter test signal. Our light source consists of a CW blinding laser and a pulsed laser (with about $16~\nano\second$ pulse width), combined on a fiber-optic beamsplitter (BS). The energy of laser pulse can be adjusted by the variable attenuator (OZ Optics DA-100). A power meter is used for monitoring the laser output power. A function generator produces trigger pulses to synchronize the laser source and signal recordings. The signal from the TES is digitized by a data acquisition module (DAQ) and analyzed on a computer (PC). The DAQ is a 16-bit, $125~\mega\hertz$ sampling rate analog-to-digital converter (AlazarTech ATS660) mounted on a peripheral component interconnect (PCI) bus of the PC. This DAQ allows measuring signals of millivolt level. Typical single-photon responses are shown in \cref{fig:oscillograms}(a). The peak voltage value during $5~\micro\second$ following the application of the optical pulse is assumed to be the amplitude of the detector response $V_\text{max}$.


Next, we investigate two potentially exploitable vulnerabilities of the TES detector. 


\emph{Wavelength-dependent response.} TES amplitude output voltage $V_\text{max}$ is inherently proportional to the energy of photons absorbed, and sensitive to a wide range of wavelengths. In principle, $N$ photons with a wavelength $N\lambda$ arriving simultaneously have the same combined energy E as one photon with the wavelength $\lambda$. This can be seen from the relation $E = N h c/\lambda$, where $h$ is Planck's constant and $c$ is the speed of light in vacuum. Thus TES would produce the same output in these two cases \cite{rosenberg2005,joshi2014,hattori2019}. 

\begin{figure}
	\includegraphics{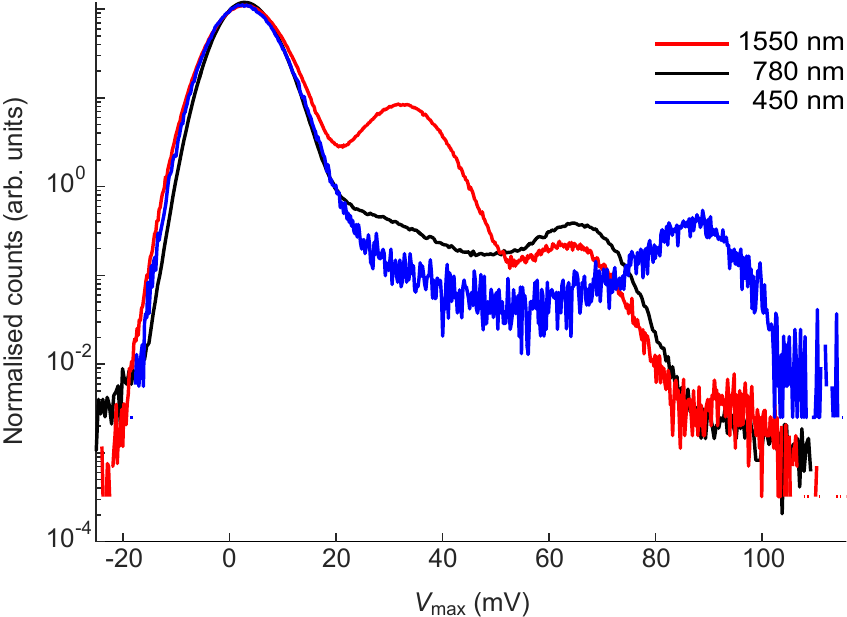}
	\caption{Histogram of TES output voltage under weak-coherent laser illumination at $1550~\nano\meter$ (red), $780~\nano\meter$ (black), and $450~\nano\meter$ (blue). The leftmost peak represents zero-photon detection. Subsequent peaks to the right represent higher photon number detections. These peaks appear at the voltage level proportional to the energy of the photons.}
	\label{fig:results-wavelength}
\end{figure}

We illustrate this fact with a simple experiment that shows how an attacker Eve could fake a single-photon detection result by sending multiple photons with proportionally lower photon energy. We send weak-coherent signals from several lasers of different wavelengths through the input fiber of the TES. We then record the voltage response's amplitude $V_\text{max}$ from the TES. The histogram in \cref{fig:results-wavelength} shows that the response signal of single-photon detection from a $450~\nano\meter$ photon is overlapped with two-photons detection from $780~\nano\meter$ and three-photons detection from $1550~\nano\meter$ photons. This shows that an expected photon number readout from the TES could be faked by multiple photons with a proportionally longer wavelength. It shows that the photon number measurement results from the TES alone cannot be used to characterize the photon number distribution of photon signal through an untrusted channel, such as the quantum channel, where the adversary could intercept and replace the signal with photons of arbitrary wavelength. Thus, any QKD scheme using photon number distribution from TES to monitor Eve's activity in the quantum channel is vulnerable to this wavelength-dependent attack \cite{xu2010e}. A narrow-band wavelength filter should prevent this attack. However, the characterization of the filter's performance against exploitable wavelengths is needed. 

\emph{Blinding attack.} In a blinding attack on QKD receiver, Eve turns the QKD detectors insensitive to single photons (blinded), but able to produce the expected detection output results when experiencing a bright-light pulse. This type of attack has been demonstrated in various single-photon detectors \cite{lydersen2010a,lydersen2010b,lydersen2011,lydersen2011c,gerhardt2011,huang2016}. 

In the ideal condition, the TES operates at the transition edge between superconductivity and normal resistive state. In this region, a small change of energy such as single-photon absorption could induce a measurable change in the output voltage proportional to the energy absorbed. By setting a voltage threshold level for each input photon energy, one could discriminate the number of absorbed photons. From the known characteristic of TES \cite{irwin2005} at a slightly higher temperature than the operational regime, it could produce the same voltage output level when absorbing much higher energy that can be delivered by a bright laser pulse. In this section, we experimentally demonstrate this behavior. 

\begin{figure}
	\includegraphics{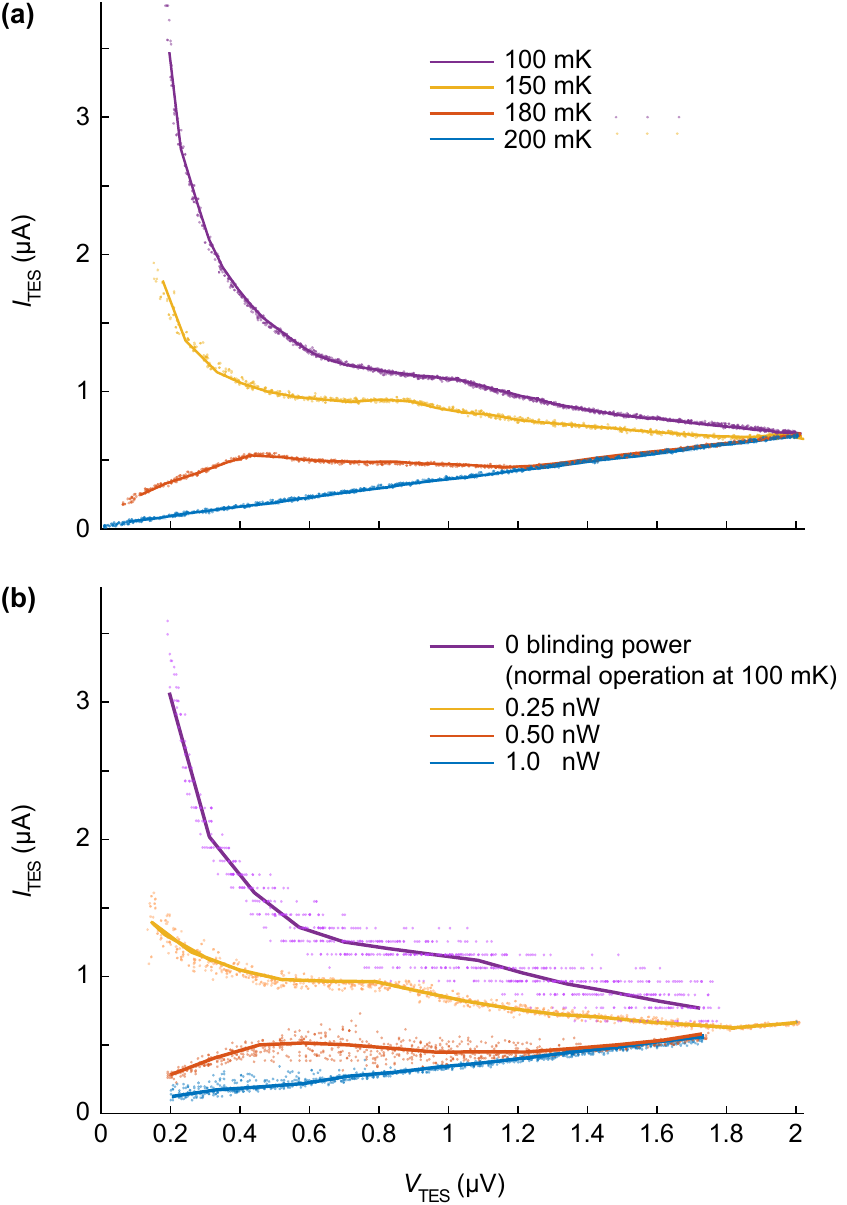}
	\caption{I--V curves of the TES. (b)~The characteristics of the system at $100~\milli\kelvin$ under bright laser illumination closely resemble (a)~the characteristics at different heat-bath temperatures. This confirms Eve's ability to control TES's temperature using bright light through the input fiber. Dots are measurement results while a solid line is their bin-averaging.}
	\label{fig:results-IV}
\end{figure}

We first investigate the behavior of TES when its temperature is increased beyond the designed transition-edge region. We set the TES to the operating temperature of $100~\milli\kelvin$. We record the current-voltage (I--V) characteristic curves of the TES at different temperatures \cite{fukuda2009}. These characteristic curves, shown in \cref{fig:results-IV}(a), will be used as a reference for the following experiments. At low temperature ($100~\milli\kelvin$), $I_{\text{TES}}$ is roughly inverse proportional on $V_\text{TES}$. As the temperature increases, $I_{\text{TES}}$ becomes lower. Once the device reaches its critical temperature of $\approx 180~\milli\kelvin$, $I_{\text{TES}}$ becomes directly proportional on $V_\text{TES}$ as the TES becomes a normal resistor.

We now demonstrate the ability of Eve to control the temperature using bright light. A CW laser at $1550~\nano\meter$ is coupled through the input fiber of TES. \cref{fig:results-IV}(b) shows that the I--V characteristics at different temperature of the device under test can be replicated. This shows that an adversary could arbitrarily control the temperature of TES using bright CW laser.

\begin{figure}
	\includegraphics{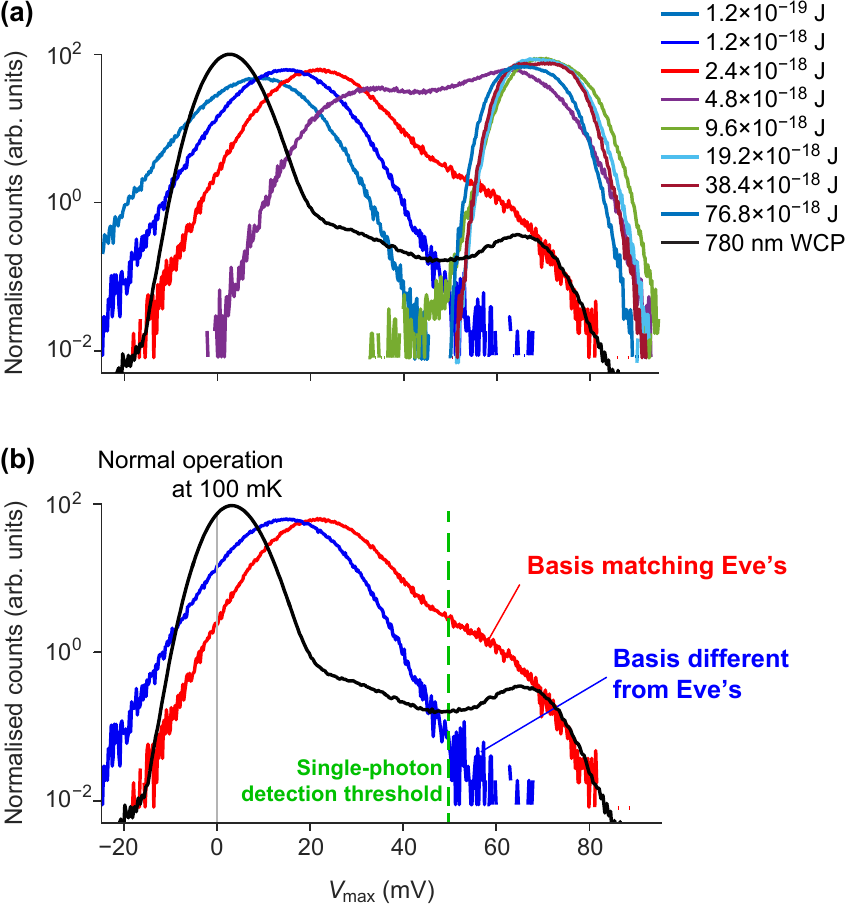}
	\caption{Detector response to the faked-state attack. For comparison, the black curve shows the normal response to a weak coherent pulse (WCP) attenuated to a single-photon level, containing the zero-photon response (left peak) and the one-photon response (right peak). (a)~Fake histogram of output voltage at different faked-state pulse energies. The detector is blinded with $0.25~\nano\watt$ CW light. (b)~An attack model on a BB84 QKD system with TES as a detector. The threshold (green vertical dashed line) marks the minimum TES voltage output that the system in our model would register as a detection. The fake response is shown for two cases where Bob and Eve pick the same (red) and different (blue) measurement bases under fake pulsed signal of $2.4\times10^{-18}~\joule$ pulse energy.}
	\label{fig:results-fake}
\end{figure}

For the faked-state attack, the appropriate blinding laser power is one that puts the response at the threshold between the transition-edge regime and the normal resistor regime.  In this region, the TES is `blinded' from single-photon input as the change of voltage produced by an additional absorption is minimal. At the same time, the system in this condition could produce the same voltage level as the system at normal operating temperature when absorbing a bright laser pulse. The histogram of faked-state results with different peak power is shown in \cref{fig:results-fake}(a) and typical oscillograms in \cref{fig:oscillograms}(b). Here, the fake signals are laser pulses with $16~\nano\second$ width and $100~\kilo\hertz$ repetition rate. The detector response exhibits a strong superlinearity \cite{lydersen2011b} between Eve's pulse energies of $1.2$ to $9.6 \times 10^{-18}~\joule$, which is a potential security loophole. I.e.,\ the voltage response of TES can be controlled by Eve who has access to the input channel. She can choose a bright laser power such that the voltage output represents a `photon number' of her choice. The physics of the detector in this regime is not clear to us and needs to be investigated further.


\emph{Attack model.} To emphasize the threat of vulnerability found in the previous section, we model a faked-state attack  \cite{lydersen2010a}  on a Bennett-Brassard 1984 (BB84) \cite{bennett1984} QKD system, assuming it uses the TES under test as its detectors. We assume here that the wavelength of the signal used by Alice and Bob is $780~\nano\meter$. In this attack model, the adversary Eve intercepts each signal from Alice and measures it in a random basis. She then reproduces a bright fake signal identical to her detection result and sends it to Bob. Here, she also sends a CW blinding laser power set to $0.25~\nano\watt$ and sets her fake pulsed signal at $2.4\times10^{-18}~\joule$ pulse energy, both at $1550~\nano\meter$. In case of Bob's measurement basis choice being different from that of Eve, the power of the fake signal would be split equally between Bob's detectors (we assumed here Bob's basis choice modulator is wavelength-independent). As shown in \cref{fig:results-fake}(b), most of the response signal from TES would fall below the single-photon detection threshold, thus remain unregistered. However, if their basis choices matched, sometimes the signal will be registered. This attack condition causes extra detection loss in Bob. Eve could hide this loss from Alice and Bob if the original quantum channel loss between Alice and Bob is lower than the detection loss induced by Eve's attack. When the basis of measurement between Eve and Bob are different, half of the registered detection events would cause an error in the key. This can be seen in the portion of the blue histogram to the right of the single-photon threshold (green line) in \cref{fig:results-fake}(b). With this estimated detection probability and error rate, the quantum bit error rate of the attack could be calculated. Our calculation shows that this attack on a QKD system with the TES under test and the specific parameters assumed above would induce $7.4\%$ quantum bit error rate (QBER). This QBER is lower than the $11\%$ abort threshold of the BB84 protocol \cite{gottesman2004}, thus the security of the key could be compromised. 

This shows a possible vulnerability of a QKD system with TES as a single photon threshold detector. A more general attack on a QKD scheme with TES as a photon number resolving detector, as well as attack on other QKD protocols such as coherent-one-way (COW) \cite{lydersen2011} can also be considered.



In conclusion, we have experimentally demonstrated two possible security vulnerabilities of TES as a photon detector. In this study, we have illustrated the ability of Eve to fake photon-number results in TES using different wavelengths. We have also shown that the characteristics of TES could be altered by a bright CW laser, and photon-number detection results could be faked using laser pulses with appropriate peak power. Using this result, we model an attack on a BB84-QKD system with TES as a detector and show that Eve could perform the intercept-and-resend attack while inducing as low as $7.4\%$ error rate, under certain specific assumptions. Since the TES under test has a misalignment of its input coupling, which limits its detection efficiency, we speculate that an attack on a higher-efficiency TES with better energy resolution might yield a better result for Eve. Understanding a physical model of the TES under attack can be a topic of a future study. Countermeasures to such attacks will need to be considered in the future when TESes begin getting employed in secure quantum communication schemes.

\medskip
This research was funded by the Ministry of Education and Science of Russia (program NTI center for quantum communications) and NSERC of Canada. P.C.\ was supported by the Thai DPST scholarship. J.Z.\ was supported in part by National Key R\&D Program of China under grant 2017YFA0304003 and in part by the National Natural Science Foundation of China under grants U1731119, U1831202, and U1931123. A.H.\ was supported by the National Natural Science Foundation of China (grant 6201101369). V.M.\ was supported by the Key program of special development funds of Zhangjiang national innovation demonstration zone (grant ZJ2018-ZD-009) and the Russian Science Foundation (grant 21-42-00040). H.Q.\ was sponsored by Shanghai Pujiang Program.

\emph{Author contributions:} P.C.,\ J.Z.,\ A.H.\ and H.Q.\ performed the experiment. P.C.\ and J.Z.\ developed the attack model and wrote the paper with input from all authors. V.M.\ and S.-c.S.\ supervised the study.

\section*{Data availability}
The data that support the findings of this study are available from the corresponding author upon reasonable request.

\bibliography{library}

\end{document}